\documentclass[12pt]{article}
\usepackage{graphicx}
\usepackage{xspace}
\usepackage[utf8x]{inputenc}

\newcommand\pubnumber{}
\newcommand\pubdate{\today}

\def\firenze{Università degli Studi di Firenze and INFN - Sezione di Firenze, \\Sesto Fiorentino, I-50019 Firenze, ITALY}
\def\support{\footnote{Contents presented on behalf of the LHCb, CMS, and ATLAS Collaborations}}

\def\Title#1{\begin{center} {\Large #1 } \end{center}}
\def\Author#1{\begin{center}{ \sc #1} \end{center}}
\def\Address#1{\begin{center}{ \it #1} \end{center}}

\newcommand\pubblock{\rightline{\begin{tabular}{l} \pubnumber\\
         \pubdate  \end{tabular}}}
\newenvironment{Abstract}{\begin{quotation}  }{\end{quotation}}
\newenvironment{Presented}{\begin{quotation} \begin{center} 
             PRESENTED AT\end{center}\bigskip 
      \begin{center}\begin{large}}{\end{large}\end{center} \end{quotation}}

\def\beq{\begin{equation}}
\def\eeq#1{\label{#1}\end{equation}}
\def\eeqn{\end{equation}}

\def\beqa{\begin{eqnarray}}
\def\eeqa#1{\label{#1}\end{eqnarray}}
\def\eeqan{\end{eqnarray}}

\let\bar=\overbar

\def\Dslash{\not{\hbox{\kern-4pt $D$}}}
\def\dslash{\not{\hbox{\kern-2pt $\del$}}}

\def\msb{{\bar{\ssstyle M \kern -1pt S}}}

\def\Bc{\ensuremath{B_c^+}\xspace}
\def\Bs{\ensuremath{B_s^0}\xspace}

\def\jpsi{\ensuremath{J\!/\!\psi}\xspace}

\def\invfb{\ensuremath{\mathrm{fb^{-1}}}\xspace}

\def\lhcb{\ensuremath{\mathrm{LHCb}}\xspace}
\def\BcJpsiMuNuX{\ensuremath{\Bc \to \jpsi\mu^+\nu_{\mu} X}\xspace}

\begin{document}
\begin{titlepage}
\pubblock

\vfill
\Title{Properties and Decays of the \Bc meson}
\vfill
\Author{ Lucio Anderlini\support}
\Address{\firenze}
\vfill
\begin{Abstract}
  Recent studies of properties and decays of the \Bc meson by the LHC experiments are presented.
  Mass and lifetime measurements are discussed and some of the many new observed decays are reported.
\end{Abstract}
\vfill
\begin{Presented}
2014 Flavour Physics and CP Violation (FPCP-2014)\\
Marseille, France, May  25--30, 2014
\end{Presented}
\vfill
\end{titlepage}
\def\thefootnote{\fnsymbol{footnote}}
\setcounter{footnote}{0}

\section{Introduction}
\begin{figure}[b]
  \centering
  \includegraphics[width=0.9\textwidth]{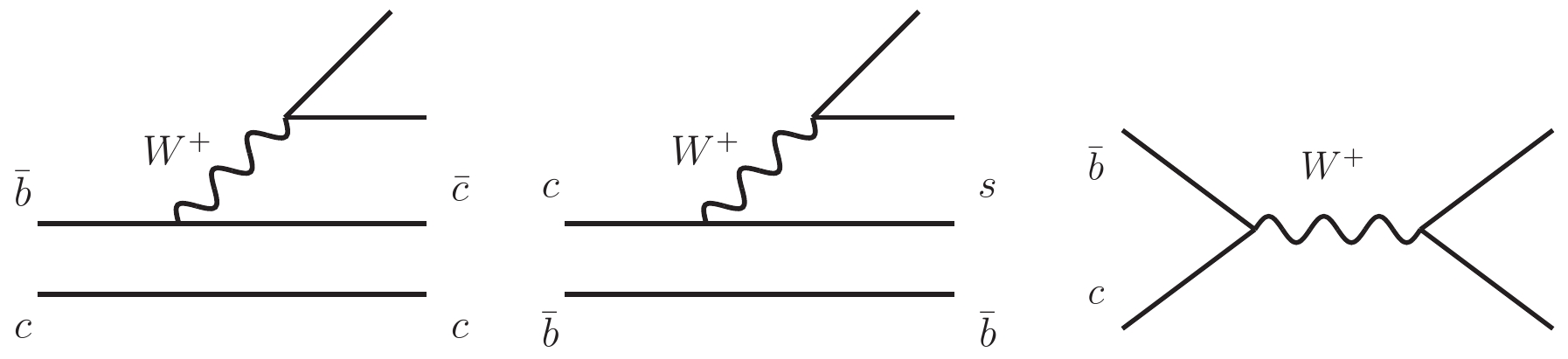}
  \caption{Feynman diagram of \Bc decays with the $c$-quark as spectator on the left, the $b$-quark spectator in the central figure, and through weak annihilation on the right.}

  \label{fig:BcDecays}
\end{figure}

The \Bc meson is the ground-state of a $(\bar b, c)$ bound system.
Being composed of two different heavy quarks it is a unique state in the Standard Model, offering an excellent laboratory to
test both QCD and weak interaction.
Indeed, being an open-favoured state, \Bc cannot decay through strong or electromagnetic interactions making weak interaction the only possible decay mechanism.
Theoretical predictions  indicate that 70\% of the total decay width of the \Bc meson is due to $c$-quark decays, 20\% to $b$-quark
decays and 10\% to weak annihilation \cite{Theoretical}. Diagrams of the three categories are shown in Figure \ref{fig:BcDecays}.

Experimentally, the \Bc was discovered at Tevatron in 1998 \cite{CDFDiscovery}, where
the CDF and D0 Collaborations observed three decays of the \Bc meson: the non-leptonic $\Bc \to \jpsi \pi^+$ decay
and the semileptonic decays $\Bc \to \jpsi \mu^+ \nu$ and $\Bc \to \jpsi e^+ \nu$.

The LHC experiments contributed significantly to improve this picture by observing a number of new decays.
Many of the observed decays have a $\jpsi$ in the final state, plus several mesons, 
for example LHCb and CMS reported the observation 
of the decay $\Bc \to \jpsi \pi^+ \pi^- \pi^+$ \cite{RcuCMS,Bc2JpsiPiPiPi}. 
LHCb also reported the observation of the decays $\Bc \to \jpsi K^+K^-\pi^+$ and $\Bc \to \jpsi 3\pi^+2\pi^-$ 
\cite{Bc2Jpsi5Pi,Bc2JpsiKKPi}.
    
Other important observations reported by LHCb 
include the decay channels $\Bc \to \psi(2S) \pi^+$ \cite{Bc2Psi2sPi}, 
 $\Bc \to \jpsi D_s^+$, and $\Bc \to \jpsi D_s^{*+}$ \cite{Bc2JpsiDs}.

The decay $\Bc \to \Bs \pi^+$, observed by LHCb, represents the first 
observation of a \Bc decay due to a $c\to s$ transition \cite{Bc2BsPi}, and is covered in Section \ref{sec:Bc2Bs}.
Production measurements are covered in Section \ref{sec:production}, while mass and lifetime measurements
are described in Sections \ref{sec:mass} and \ref{sec:lifetime}, respectively.

\section{Observation of $\Bc \to \Bs \pi^+$ decays} \label{sec:Bc2Bs}
    LHCb reported the observation of the $\Bc \to \Bs \pi^+$ decay 
     analysing the 
    $pp$-collision data collected in 2011 at a center-of-mass energy of 7 TeV, and in 2012 at a center-of-mass energy of 8 TeV,
    corresponding to an integrated luminosity of 1 fb$^{-1}$ and 2 fb$^{-1}$, respectively.
    
\begin{figure}[htb]
      \centering
      \includegraphics[width=0.45\textwidth]{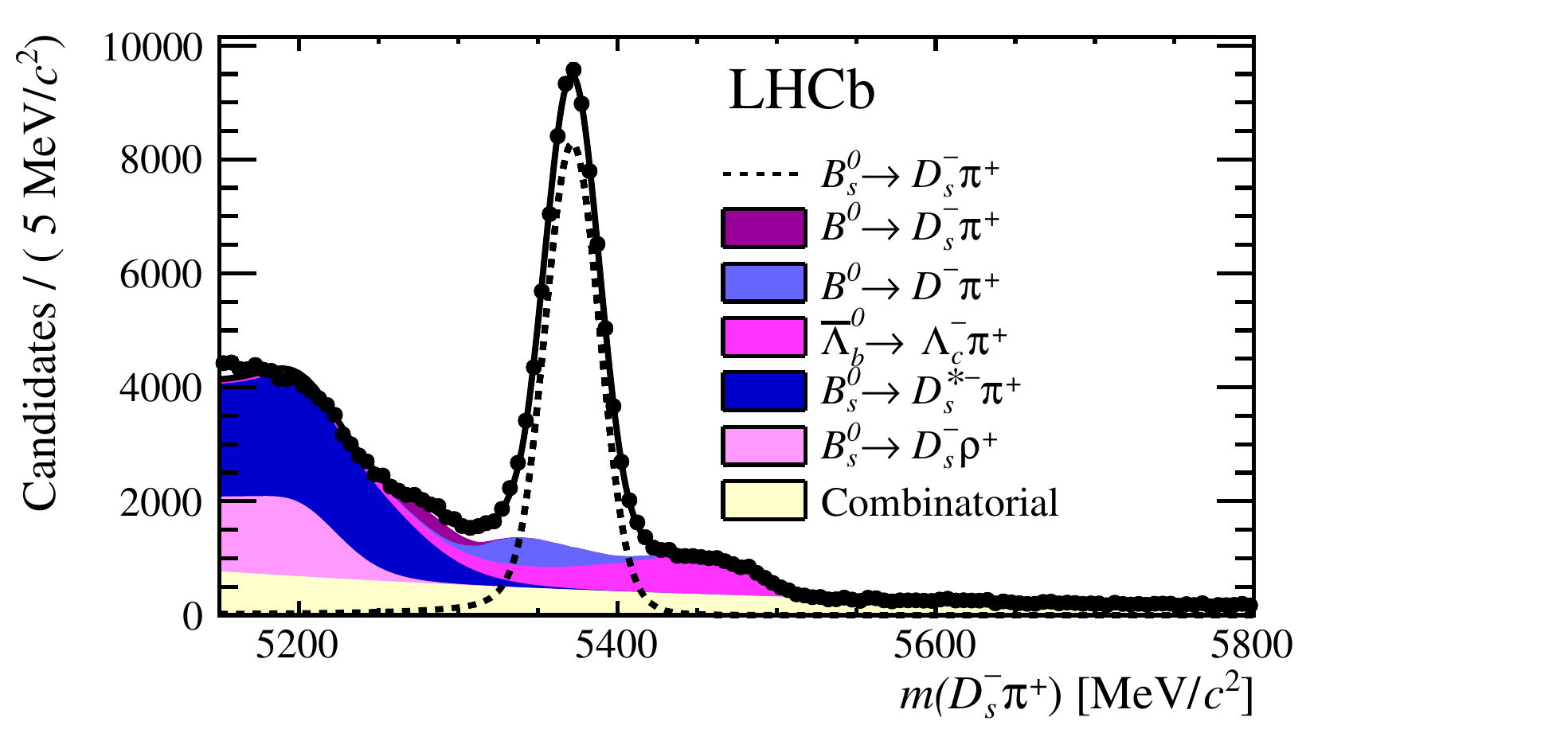}
      \includegraphics[width=0.35\textwidth]{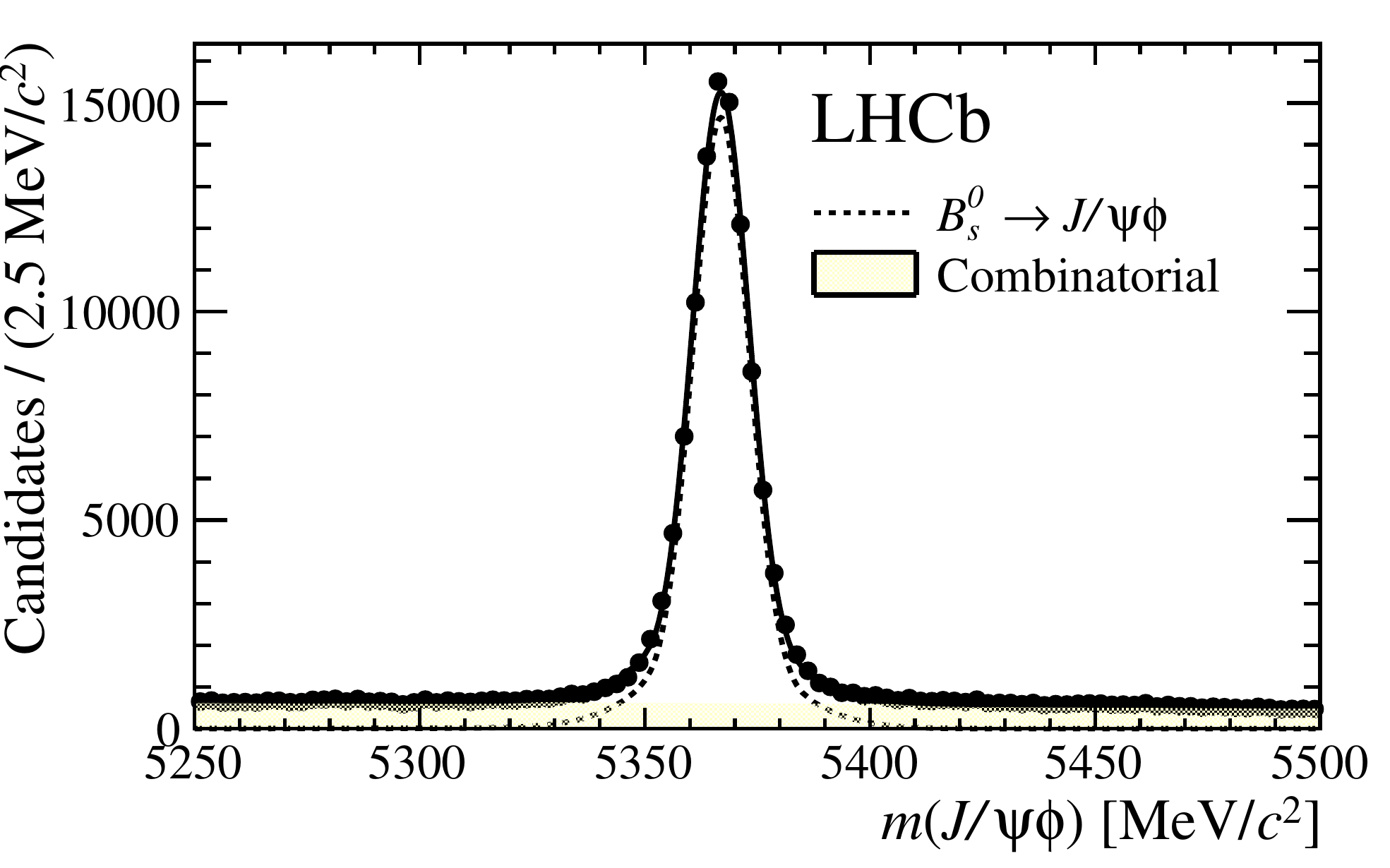}
      \caption{\label{fig:BsMass} \Bs mass distribution for the candidates reconstructed 
               as $\Bs \to D_s^- \pi^+$ (left) and $\Bs \to \jpsi \phi$ (right).
      }
    \end{figure}

    Predictions for the branching fraction $\mathcal{B}(\Bc \to \Bs\pi^+)$ span a large range
    between 2.5\% and 16.4\% (see Ref. \cite{Bc2BsTheory}, and references therein). LHCb has performed a search starting from two samples of 
    fully reconstructed \Bs mesons decaying to $\Bs \to D_s^- \pi^+$ and $\Bs \to \jpsi\phi$.
    The distributions of the invariant mass of the \Bs candidates, as reconstructed in the 
    two final states, are shown in Figure \ref{fig:BsMass}.

\begin{figure}[htb]
      \centering
      \includegraphics[width=0.49\textwidth]{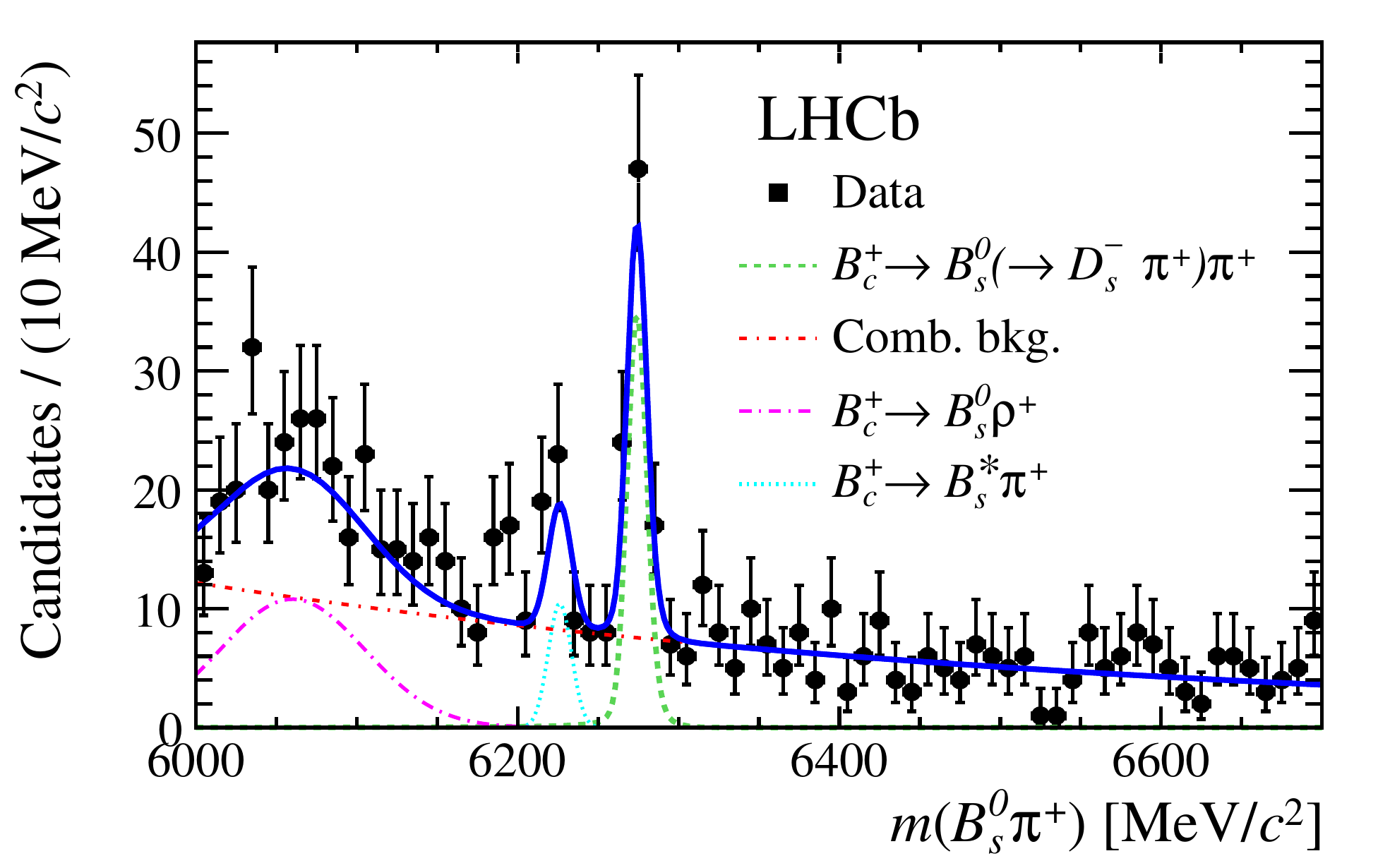}
      \includegraphics[width=0.49\textwidth]{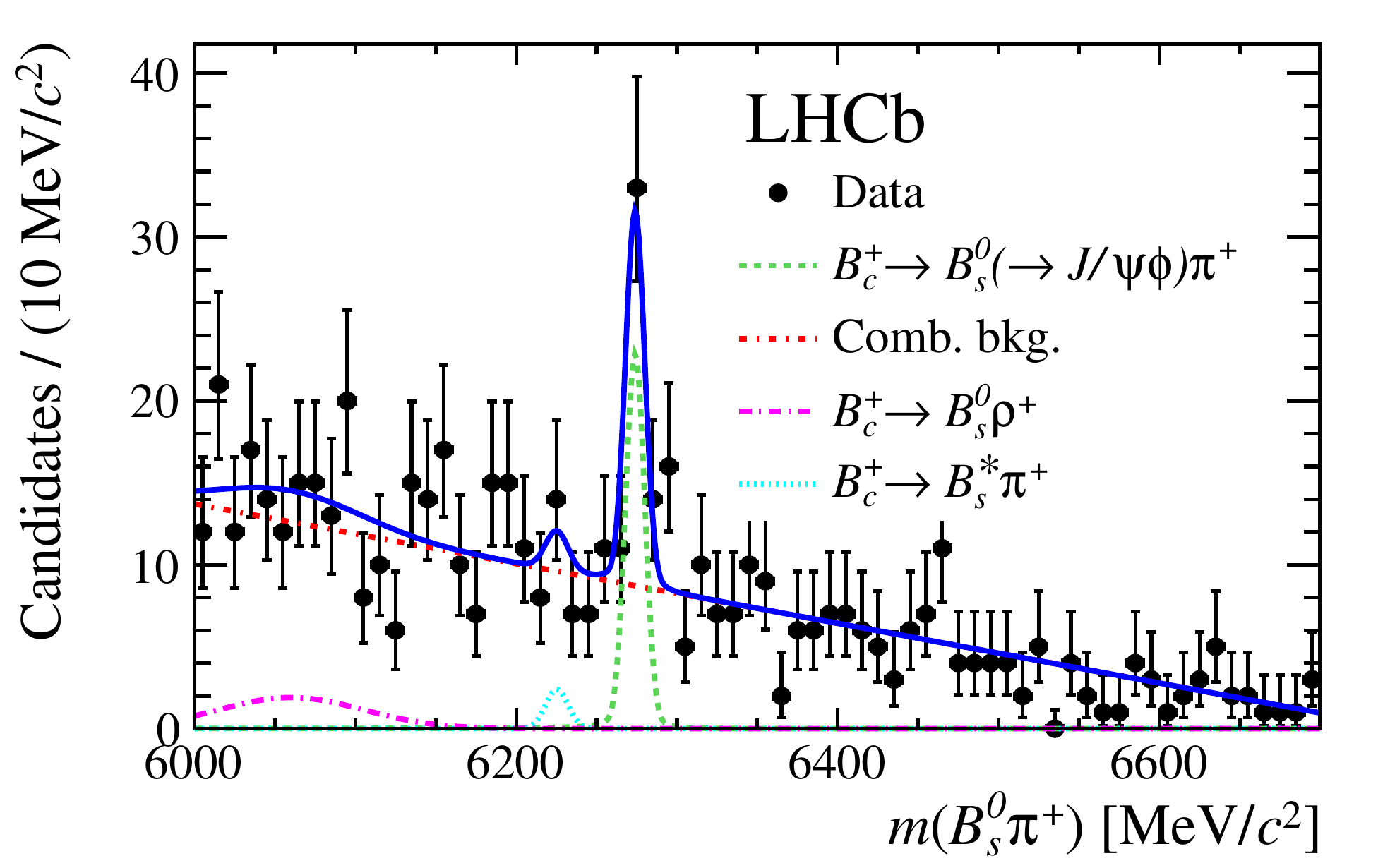}
      \caption{\label{fig:BcMass} Invariant mass for the $\Bs\pi^+$ combinations with the \Bs
               reconstructed as $\Bs \to D_s^- \pi^+$ (left) and $\Bs \to \jpsi \phi$ (right).
      }
    \end{figure}

    $ 73\, 700 \pm 500$ $\Bs \to D_s^-\pi^+$ and
    $103\, 760 \pm 380$ $\Bs \to \jpsi\phi$ candidates are observed and combined to a charged pion 
    to create \Bc candidates.
    The distributions of the invariant mass of such candidates 
    are shown in Figure \ref{fig:BcMass}.
    
    The fitted signal yield for $\Bc \to \Bs(\to D_s^-\pi^+)\pi^+$ decays is $64 \pm 10$ 
    corresponding to a statistical significance of $7.7\sigma$;
    for $\Bc \to \Bs(\to \jpsi\phi)\pi^+$, $35 \pm 8$ signal candidates are observed, corresponding 
    to a statistical significance of $6.1\sigma$. 
    
    The \Bs and \Bc yields are corrected for the relative detection efficiencies, to obtain 
    the efficiency-corrected ratios of $\Bc\to\Bs\pi^+$ over $B_s^0$ yields,
    \begin{equation}
      \left(2.54 \pm 0.40 (\mathrm{stat})\ ^{+0.23}_{-0.17}(\mathrm{syst})\right)\times 10^{-3},
      \quad \mathrm{and} \quad
      \left(2.20 \pm 0.49 (\mathrm{stat})\pm 0.23 (\mathrm{syst})\right)\times 10^{-3}.
    \end{equation}
    for \Bs reconstructed as $D_s^-\pi^+$ and $\jpsi\phi$ respectively.
    The systematic uncertainty is dominated by the uncertainty on the lifetime of the \Bc meson
    which results into an uncertainty on the selection efficiency of criteria based on \Bc flight distance.
    Such contribution is correlated between the two \Bs reconstruction channels,
    and is therefore taken into account separately when combining the results above 
    to give the ratio of production rates
    multiplied with the branching fraction
    \begin{equation}
      \frac{\sigma(\Bc)}{\sigma(\Bs)} \times \mathcal B(\Bc \to \Bs \pi^+) = 
        \left(2.37 \pm 0.31 (\mathrm{stat}) \pm 0.11 (\mathrm{syst}) ^{+0.17}_{-0.13} (\tau_{\Bc}) )\right).
    \end{equation}
    
    Assuming a value for ${\sigma(\Bc)}/{\sigma(\Bs)}$ of 0.2 \cite{Bc2BsPi}, one would obtain
    a branching ratio $\mathcal B(\Bc \to \Bs \pi^+)$ of about 10\%, the highest branching 
    fraction ever observed for a $b$-hadron weak decay.

\section{Production measurements}\label{sec:production}
  The production mechanism of the \Bc meson at hadronic machines 
  is expected to be quite different from the production mechanism of hadrons containing a 
  single heavy quark.
  According to Ref. \cite{Theoretical2}, the probability of a fragmentation of a $b$ quark towards 
  a \Bc state is about 11 order of magnitude lower
  than a fragmentation towards $B^0$ or $B^+$ states.

  This means that other production mechanisms, mainly $gg$ fusion processes, dominate the production of such states.
  Perturbative QCD calculations at the fourth order in $\alpha_s$ lead to expectations for the \Bc production 
  cross-section at LHC between 0.4 and 0.9 $\mu$b, roughly corresponding to $10^{-3} \sigma(B_{u,d})$, ten times larger than
  the production cross-section at Tevatron \cite{Theoretical}.
  
  Since no absolute branching fraction has never been measured for \Bc states, CMS and LHCb have used $B^+ \to \jpsi K^+$ as normalization channel, measuring the ratio
  \begin{equation}
    R_{c/u} = \frac{\sigma(\Bc) \times \mathcal B (\Bc \to \jpsi \pi^+)}{\sigma(B^+) \times \mathcal B (B^+ \to \jpsi K^+)}
            = \frac{N(\Bc \to \jpsi \pi^+)}{\epsilon_{tot}^{c}} \frac{\epsilon_{tot}^u}{N(B^+ \to \jpsi K^+)}
  \end{equation}
  where $N(\Bc \to \jpsi \pi^+)$ and  $N(B^+ \to \jpsi K^+)$ represents the number of reconstructed \Bc and $B^+$ decays, respectively,
  and are corrected for the total selection efficiencies $\epsilon_{tot}^{c}$ and $\epsilon_{tot}^{u}$, respectively.
  The choice of the normalization channel $B^+ \to \jpsi K^+$ makes many of the uncertainties to cancel in the ratio.

  Experimental values for $R_{c/u}$ were measured by CMS and LHCb in two different kinematical regions.

  \begin{equation}
    R_{c/u}^{\mathrm{CMS}} = \left(0.48 \pm 0.05 (\mathrm{stat}) \pm 0.04 (\mathrm{syst}) ^{+0.05}_{-0.03} ({\tau_{\Bc}})\right) \times 10^{-2}
  \end{equation}
  with $p_T(\Bc) > 15~\mathrm{GeV}/c$, $|y| < 1.6$, $\sqrt{s} = 7~\mathrm{TeV}$, $\mathcal L = 5.1~\invfb$ \cite{RcuCMS}.

  \begin{equation}
    R_{c/u}^{\mathrm{LHCb}} = \left(0.68 \pm 0.10 (\mathrm{stat}) \pm 0.03 (\mathrm{syst}) \pm 0.05({\tau_{\Bc}})\right) \times 10^{-2}
  \end{equation}
  with $p_T(\Bc) > 4~\mathrm{GeV}/c$, $2.5 < \eta < 1.6$, $\sqrt{s} = 7~\mathrm{TeV}$, $\mathcal L = 0.37~\invfb$ \cite{RcuLHCb}.

  The third error is due to the uncertainty on the lifetime of the \Bc meson which reflects into an uncertainty on the selection efficiency
  of criteria correlated to the \Bc flight distance.

\section{Mass measurements} \label{sec:mass}
\begin{figure}
    \centering
    \includegraphics[width=0.36\textwidth,trim = 0 0 0 0]{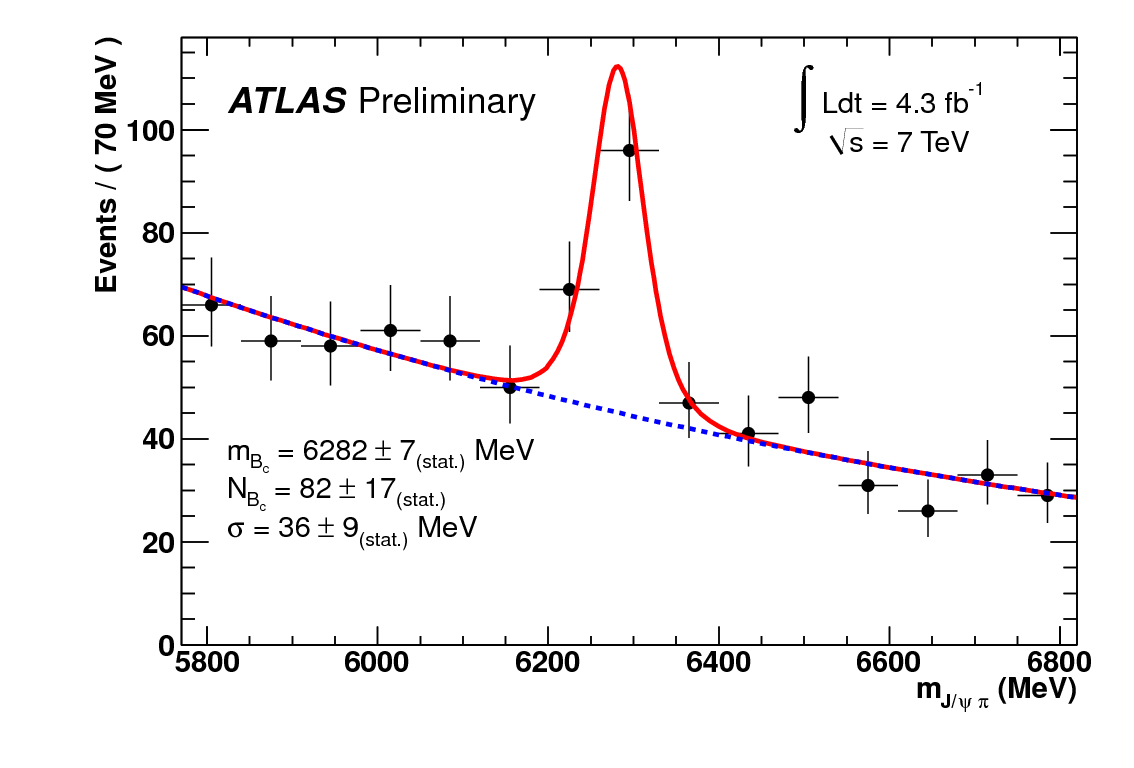}
    \includegraphics[width=0.25\textwidth,trim = 0 0 0 0]{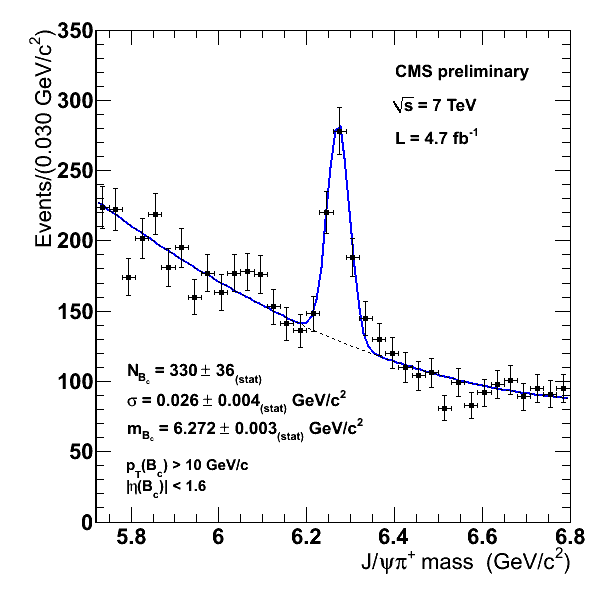}
    \includegraphics[width=0.35\textwidth,trim = 0 30pt 0 0]{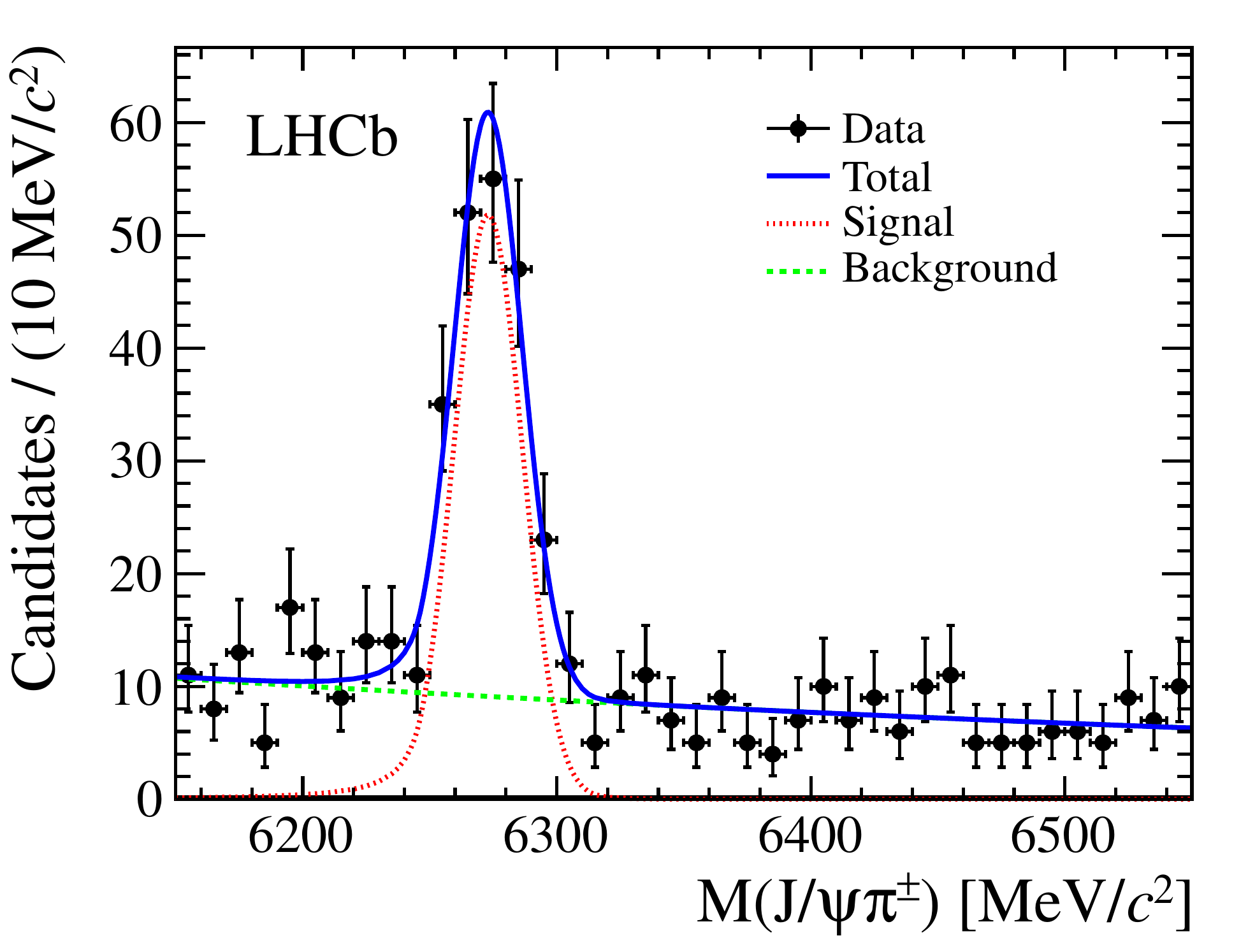}
    \caption{\label{fig:MassJpsiPi} Invariant mass of the selected $\jpsi\pi^+$ combinations as used in the \Bc mass measurement
              by the ATLAS (left), CMS (center), and LHCb (right) Collaboration.}
  \end{figure}

  The ATLAS and CMS Collaborations reported a preliminary measurement of the \Bc mass studying the decay channel 
  $\Bc \to \jpsi\pi^+$.
  ATLAS performed the measurement using $82 \pm 17$ candidates, shown in Figure \ref{fig:MassJpsiPi}, left, 
  collected in a dataset of $pp$ collisions at $\sqrt{s} = 7$ TeV
  corresponding to an integrated luminosity of 4.3~\invfb \cite{ATLASMass}
  \begin{equation}
    M(\Bc)^{\mathrm{ATLAS (\jpsi\pi)}} = 6282 \pm 7 (\mathrm{stat})~\mathrm{MeV}/c^{2}.
  \end{equation}

  Analogously CMS performed a mass measurement on $330 \pm 17$ $\Bc \to \jpsi \pi^+$ candidates (see Figure \ref{fig:MassJpsiPi}, center)
  collected in a dataset corresponding to 4.7~\invfb
  and found \cite{RcuCMS}
  \begin{equation}
    M(\Bc)^{\mathrm{CMS(\jpsi\pi)}} = 6272 \pm 3 (\mathrm{stat})~\mathrm{MeV}/c^{2}.
  \end{equation}
  
  Using a small data sample collected in 2011 and corresponding to an integrated luminosity
  of 0.37~\invfb, LHCb reported the world best measurement of the \Bc mass studying the channel
  $\Bc \to \jpsi \pi^+$ \cite{BcProduction}: 
  \begin{equation}
    M(\Bc)^{\mathrm{LHCb(\jpsi\pi)}} = \left(6273.7 \pm 1.3 (\mathrm{stat}) \pm 1.6 (\mathrm{syst})\right)\mathrm{MeV/}c^{2}  
  \end{equation}
  The $162 \pm 18$ selected signal candidates were used to perform the mass measurement whose uncertainty 
  was already dominated by systematic uncertainty on 
  the momentum scale because of the large $Q$-value of the \Bc decay.

  Better resolution and momentum scale calibration uncertainty have been obtained two years later 
  studying the decay $\Bc \to \jpsi D_s^+$ \cite{Bc2JpsiDs}:  
  \begin{equation}
    M(\Bc)^{\mathrm{LHCb(\jpsi D_s^+)}} = \left( 6276.28 \pm 1.44 (\mathrm{stat}) \pm 0.36 (\mathrm{syst}) \right)~\mathrm{MeV/}c^{2}.
  \end{equation}
  Up to date, this is the world's most precise single measurement of the \Bc mass.
  
\begin{figure}[htb]
    \centering
    \includegraphics[width=.8\textwidth]{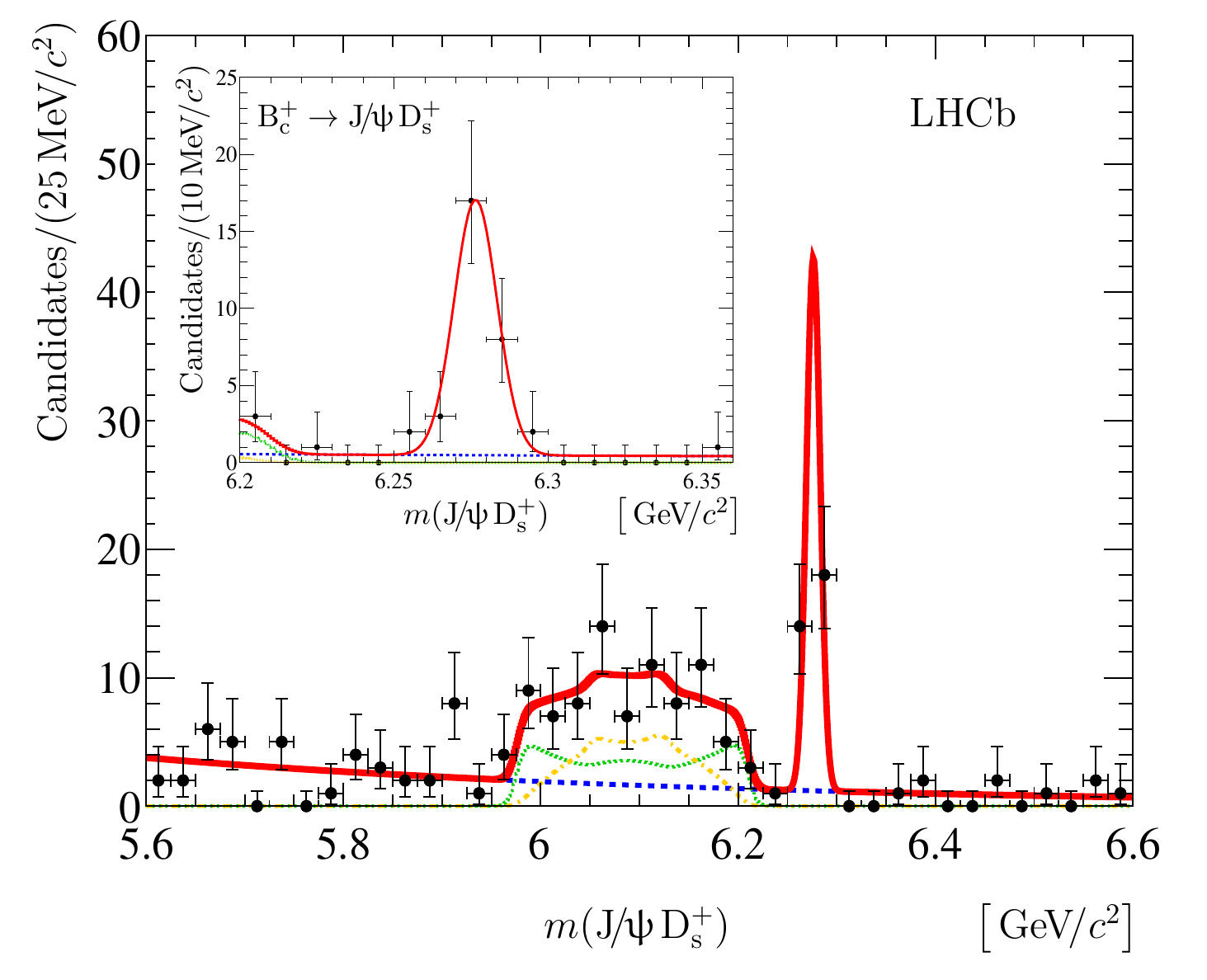}
    \caption{\label{fig:Bc2JpsiDs} \Bc mass distribution for candidates reconstructed as $\Bc \to \jpsi D_s^+$ 
             as used in the mass measurement. 
             The broader structure at lower mass is due to partially reconstructed $\Bc \to \jpsi D_s^{*+}$ decays.}
  \end{figure}

  The signal yield of $28.9 \pm 5.6$ $\Bc \to \jpsi D_s^+$ is small compared to the size of the sample
  available for $\Bc\to\jpsi\pi$, but because of the low $Q$-value
  of the decay, the experimental resolution, as shown in Figure \ref{fig:Bc2JpsiDs}, is much better and 
  therefore the statistical uncertainties are similar for the two channels.
  The systematic uncertainty is still dominated by momentum scale calibration (accounting for 0.30 MeV/$c^2$)
  on the \Bc meson mass, while other relevant contributions are the uncertainty on the $D_s^+$ mass (0.16 MeV/$c^{2}$),
  and signal modelling including simulation effects (0.11 MeV/$c^2$).
  The uncertainty on the $D_s^+$ meson mass and on the momentum scale largely cancels in the mass difference
  \begin{equation}
    m_{\Bc} - m_{D_s^+} = 4607.97 \pm 1.44 (\mathrm{stat}) \pm 0.20 (\mathrm{syst}) \ \mathrm{MeV}/c^{2}.
  \end{equation}
  
\section{Lifetime measurement with the semileptonic decay channel \BcJpsiMuNuX} \label{sec:lifetime}
  The lifetime of the \Bc meson is an important quantity for both theory and experiments.
  Many models describing heavy-quark properties can be used to predict the \Bc lifetime,
  the more precise is the knowledge of this quantity, the more theoretical models are constrained.

  Experimentally, the uncertainty on the 
  lifetime results in an uncertainty on the selection efficiency of criteria based on the 
  \Bc flight distance. Since these criteria are very powerful in rejecting background, 
  most of \Bc analyses rely on them and are affected by lifetime uncertainty.
  In most of the analyses described above, the uncertainty on the world average for the \Bc lifetime is actually a dominant systematic uncertainty, 
  so that a precise measurement of the \Bc lifetime will 
  improve the precision on most of the results reported.

  LHCb has recently measured \cite{BcLifetime} 
  the lifetime of the \Bc meson by studying the semileptonic
  decay channel \BcJpsiMuNuX, with the \jpsi reconstructed as a muon pair.  
  The large branching fraction and the very clear experimental
  signature (a 3--muon vertex),  allow this analysis to be performed 
  with a cut-based decay-time-unbiased selection, avoiding the need
  to measure the acceptance as a function of the decay time, 
  introducing the largest systematic uncertainty in most of the lifetime measurements.

  On the other hand, when studying partially reconstructed decays 
  background rejection is more difficult because the selection cannot 
  rely on a mass peak. In the case of \BcJpsiMuNuX, the invariant 
  mass of the $\jpsi\mu^+$ combination lies in a range between 
  3.2 and 6.25 GeV/$c^2$ and the shape of the distribution depends on 
  the form-factors of the decay, for which no experimental measurement is available.
  
  A correction, named $k$-factor, between the $\jpsi\mu^+$ combination rest frame and the 
  \Bc rest frame, needed to calculate the \textit{proper} decay-time and therefore the lifetime.
  The $k$-factor correction is applied on a statistical basis in bins of the mass reconstructed 
  for the $\jpsi\mu^+$ combination. The shape of the $k$-factor distribution is determined using 
  simulation and is affected by:
    (i) form-factor model of the \Bc decay;
    (ii) model of acceptance and efficiency as function of the kinematic variables;
    (iii) feed-down decays, where the final state $\jpsi\mu^+$ is reached through an intermediate state, \textit{e.g.} $\Bc \to \psi(2S)(\to \jpsi\pi^+\pi^-)\mu^+\nu_\mu$.

  Background is dominated by events in which a real \jpsi from a $b$-hadron
  decay is associated to a hadron misidentified as a muon.
  This background source, named for brevity \textit{misidentification background},
  is modeled with a data-driven technique requiring an accurate characterization of 
  the PID performance of the \lhcb detector.
  
  Other non-negligible background sources are due to decay candidates with a fake \jpsi and 
  candidates obtained combining a \jpsi with a muon, but not from the same vertex (combinatorial background).

  The \Bc lifetime is obtained through 
  a two-dimensional fit on the joint distribution of $M(\jpsi\mu^+)$ and $t_{\mathrm{ps}}$,
  the decay time reconstructed in the $\jpsi\mu^+$ rest-frame.
  
  The result,
  \begin{equation}
    \tau_{\Bc} = 509 \pm 8\ (\mathrm{stat}) \pm 12\ (\mathrm{syst})\ \mathrm{fs},
  \end{equation}
  is the world's best measurement of the \Bc lifetime, with an uncertainty
  halved with respect to the world average.
  The systematic error is dominated by the uncertainties on the background ($\pm\, 10\ \mathrm{fs}$),
  and signal ($\pm\, 5\ \mathrm{fs}$) models, where the latter includes theoretical
  uncertainties on form-factors and branching fractions of the feed-down decays.
   
  Further improvements on the precision of the lifetime measurement are expected studying the 
  hadronic decay channel $\Bc \to \jpsi \pi^+$ where systematic uncertainties
  are largely uncorrelated with those affecting the measurement presented above.

  \begin{figure}[htb]
    \includegraphics[width=0.48\textwidth]{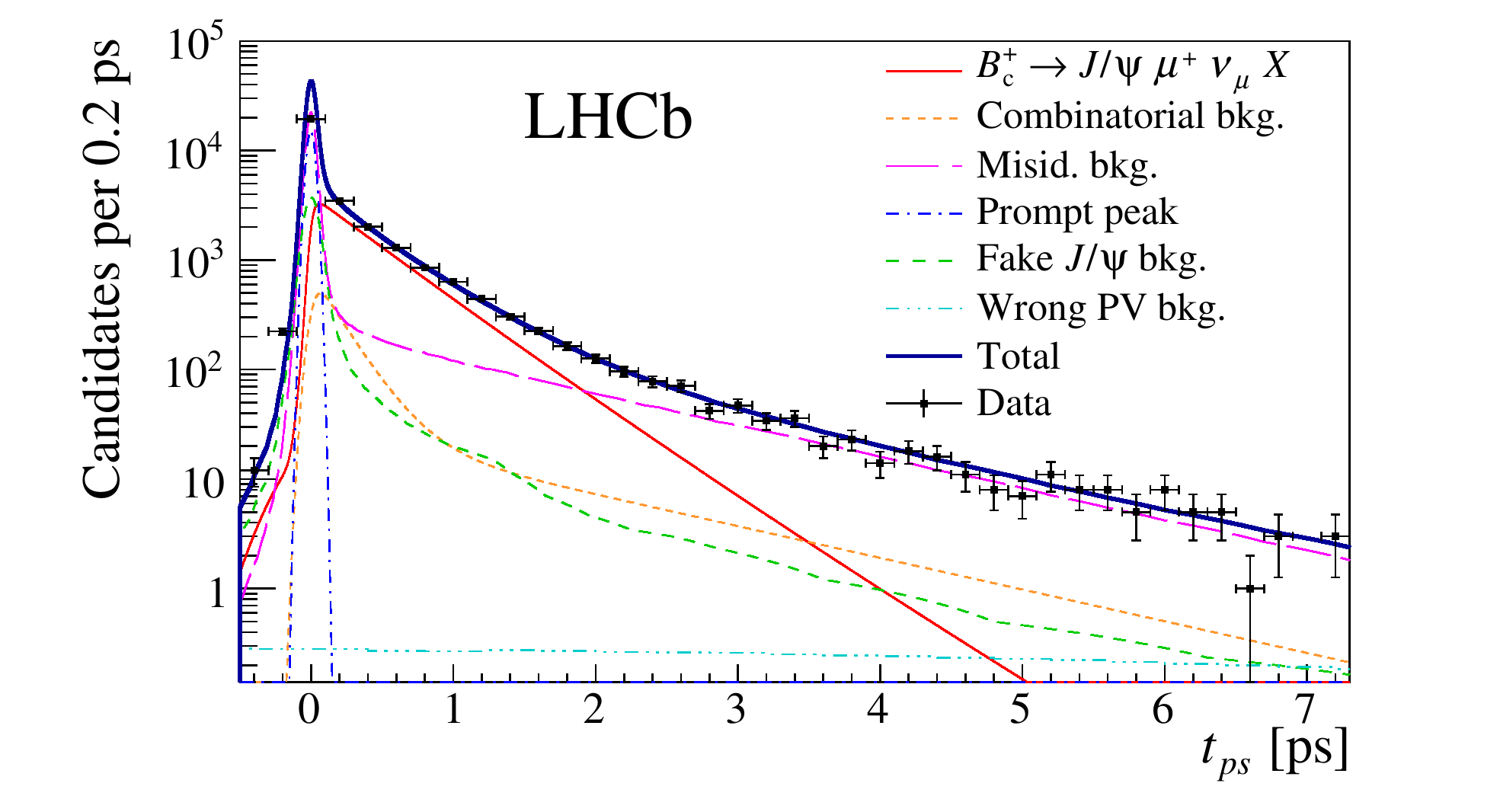}
    \includegraphics[width=0.48\textwidth]{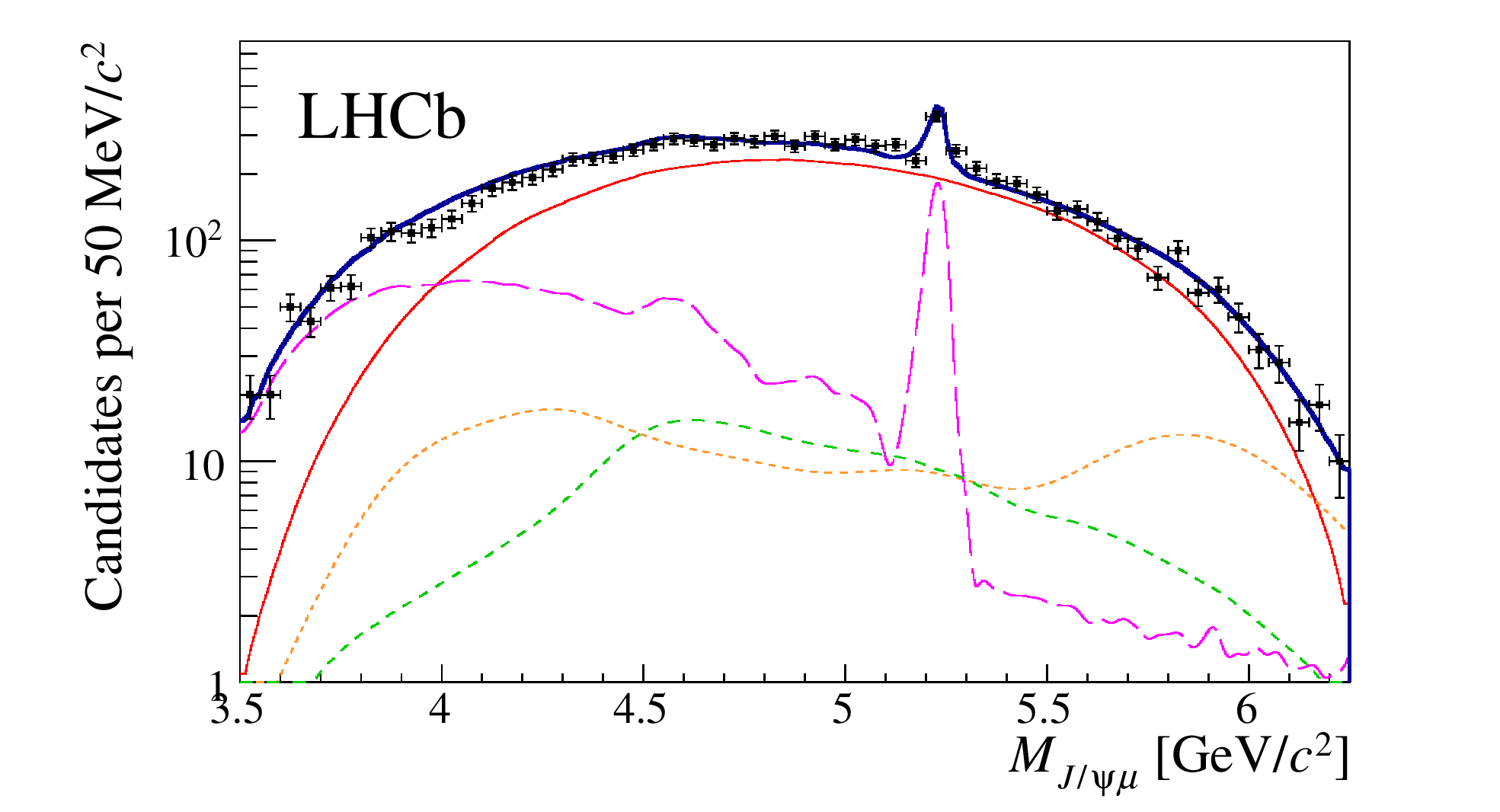}
    \caption{\label{fig:BcLifetime} Data model for the \BcJpsiMuNuX decays used in the \Bc lifetime measurement.
             On the left the model is projected on the reconstructed pseudo-proper decay time, 
             on the right on the invariant mass of the $\jpsi\mu^+$ combination.}
  \end{figure}

\section{Conclusion and outlook} \label{sec:conclusion}
  The excellent performance of the LHC and of the detectors has allowed the LHC experiments
  to reach several achievements in the field of \Bc physics.
  
  The world's best measurement of mass and lifetime were presented here together
  with the first observation of a $c\to s$ transition in a \Bc decay. 
  Several other decay channels \cite{Bc2JpsiPiPiPi,Bc2Jpsi5Pi,Bc2JpsiKKPi,Bc2Psi2sPi,Bc2JpsiK} 
  have been observed for the first time and their
  relative decay branching fraction measured.

  Between the FPCP meeting and the preparation of these proceedings new exciting results 
  have been obtained. 
  LHCb has published the measurement of the ratio of \Bc branching fraction to $\jpsi \pi^+$ and $\jpsi \mu^+$, 
  using a data sample corresponding to an integrated luminosity of 1~\invfb \cite{SLratio}.
  The ATLAS Collaboration has published the observation of an excited \Bc meson state 
  decaying to $\Bc\pi^+\pi^-$ \cite{BcStSt}.
  
  In 2015, the LHC experiments will restart data-taking of the $pp$ collisions at 13-14 TeV.
  Higher luminosity and larger \Bc production cross-section are expected and therefore
  many unobserved \Bc decay channels are expected to become accessible.
  Observed decays will be used as high-statistics control channels and 
  studied for precision measurements.





\end{document}